\documentclass{aa}
\usepackage{graphicx}
\usepackage{lscape}
\usepackage{natbib}
\usepackage{txfonts}
\bibpunct[ ]{(}{)}{;}{a}{}{,}

\def\spose#1{\hbox to 0pt{#1\hss}}
\def\lta{\mathrel{\spose{\lower 3pt\hbox{$\mathchar"218$}}
        \raise 2.0pt\hbox{$\mathchar"13C$}}}
\def\gta{\mathrel{\spose{\lower 3pt\hbox{$\mathchar"218$}}
        \raise 2.0pt\hbox{$\mathchar"13E$}}}

\begin{document}

 \title{The thermal--viscous disk instability model in the AGN context}
 \authorrunning{Hameury et al.}
 \titlerunning{The DIM in AGNs}
 \author{Jean-Marie Hameury\inst{1}, Maxime Viallet$^1$, and Jean-Pierre Lasota\inst{2,3}}
 \institute{ Observatoire de Strasbourg, CNRS/Universit\'e Louis Pasteur, 11 rue
             de l'Universit\'e, F-67000 Strasbourg, France \\
             \email{hameury@astro.u-strasbg.fr, viallet@astro.u-strasbg.fr}
 \and
             Institut d'Astrophysique de Paris, UMR 7095 CNRS, UPMC Univ Paris 06, 98bis Bd Arago,
             75014 Paris, France
             \and
              Astronomical Observatory, Jagiellonian University, ul.\ Orla 171, 30-244 Krak\'ow,
              Poland\\
             \email{lasota@iap.fr} }
 \offprints{J.M. Hameury}

\date{}

\abstract
 {Accretion disks in AGN should be subject to the same type
of instability as in cataclysmic variables (CVs) or in low-mass
X-ray binaries (LMXBs), which leads to dwarf nova and soft X-ray
transient outbursts. It has been suggested that this thermal/viscous
instability can account for the long term variability of AGNs.}
 {We test this assertion by presenting a systematic study of the application
of the disk instability model (DIM) to AGNs.}
 {We are using the
adaptative grid numerical code we have developed in the context of
CVs, enabling us to fully resolve the radial structure of the disk.}
 {We show that, because in
AGN disks the Mach numbers are very large, the heating and cooling
fronts are so narrow that they cannot be resolved by the numerical
codes that have been used until now. In addition, these fronts
propagate on time scales much shorter than the viscous time. As a
result, a sequence of heating and cooling fronts propagate back and
forth in the disk, leading only to small variations of the accretion
rate onto the black hole, with short quiescent states occurring for
very low mass transfer rates only. Truncation of the inner part of
the disk by e.g. an ADAF does not alter this result, but enables
longer quiescent states. Finally we discuss the effects of
irradiation by the central X-ray source, and show that, even for
extremely high irradiation efficiencies, outbursts are not a natural
outcome of the model.}
 {}

\maketitle

\section{Introduction}

Accretion disks are found in a large variety of astronomical
objects, from young stars to active galactic nuclei. Among these,
close binaries have deserved special attention, because they are
nearby, and vary on short timescales that enable time-dependent
studies of their light curve. In particular, a number of these
systems show large outbursts, as for example dwarf novae, which are
a subclass of cataclysmic variables in which a low-mass companion
transfers mass onto a white dwarf; these systems undergo outbursts
lasting at least a few days during which their brightness increases
by several magnitudes \citep[see e.g.][for a review]{w95}. The
outbursts are believed to be due to a thermal--viscous accretion
disk instability \citep{mm81} which arises when the disk effective
temperature becomes of order of $\lta$10$^4$K, enough for hydrogen
to become partially ionized and opacities to depend strongly on
temperature \citep[see][for a review of the model]{l01}. Similarly,
soft X-ray transients, which are a subclass of low-mass X-ray
binaries in which the compact object is either a black hole or a
neutron star also show outbursts, their amplitude being larger and
the time scales longer than for dwarf novae. The ionization
instability of the accretion disk is also thought to be cause of the
outbursts, the difference with dwarf novae being due to the
difference in the mass of the compact object (and thus in the depth
of the gravitational potential well) and to the effect of
illumination of the disk, much more important in the case of X-ray
binaries \citep[see e.g.][]{dhl01}.

It was realized long ago \citep{ls86} that the same instability
could be present in accretion disks around AGNs; it was found that,
at radii $\sim 10^{15 - 16}$ cm where the effective temperature is
indeed of a few thousand degrees, the disk should be unstable. For
the parameters of AGNs, the implied timescales are of order of $10^4
- 10^7$ yr, making impossible the direct observation of the
instability, but predicting that in many systems the disk should not
be in viscous equilibrium and that many AGNs should be in a
quiescent state \citep[see][]{sck,se}. It was also immediately
realized that, as in dwarf-novae, the character of putative AGN
outbursts strongly depends on assumptions one makes about the disk
viscosity \citep{ms90}. However, while in the case of dwarf-novae
one is guided by the \textsl{observed} outburst properties when
fixing the viscosity prescription, in the case of AGN it is not even
clear that outbursts are present as the variability of these objects
could be due just to mass-supply variations. This state of affair
gave rise to various, more or less arbitrary, prescriptions of how
viscosity varies (or not) with the state of the accretion flow
\citep{ms90,mq01,jcs04}. In addition, results of numerical
calculations of AGN outbursts were marred by the insufficient
resolution of grids used. As showed by \citet{hmdl98} low grid
resolution often leads to unreliable results.\footnote{\citet{mp06}
mischievously remark in this context that ``mathematical convergence
does not necessarily imply more accurate modeling of physical
reality". While this might be true it is clear that the lack of
convergence of a mathematical model makes it useless for physical
applications.}

The aim of the present article is the systematic analysis of the
application of the DIM in the context of AGN disks. Instabilities
other than the thermal-viscous instability may exist in AGN disks
\citep[beyond the MRI instability which is thought to be the source
of viscosity][]{bh91} and in particular the gravitational
instability that arises when self-gravity exceeds the combined
action of pressure and Coriolis forces \citep{t64,s60}; conditions
for the onset of this instability are met at large distances from
the black hole \citep[see e.g.][]{s90}. The outcome of this
instability in the AGN case is most probably the fragmentation of
the accretion disk \citep[see e.g.][]{g01,g03,r07} since in the AGN
case, the cooling time is likely to be short. \citet{db06} suggested
that the gravitational instability might instead be a source of
turbulence, which could be the case if the non-linear development of
the instability does not lead to fragmentation, not a likely outcome
in the AGN case as mentioned above. Other local or global
instabilities may arise, such as the Lightman-Eardley instability
\citep{le74}, but it is far beyond the scope of this paper to
discuss them all, and we consider parameters such as these
instabilities do not occur.

\section{Vertical disk structure}

We recall here the vertical-structure equations adapted to AGN
parameters. We consider here only the case where the viscosity $\nu$
is proportional to the gas pressure (not the total pressure, in
order to avoid the \citet{le74} instability). The vertical structure
of an $\alpha$ disk in which the viscosity $\nu$ is assumed to be
proportional to the gas pressure is given by the standard disk
equations \citep[see e.g.][and references therein]{fkr02}:
\begin{eqnarray}
\lefteqn{{dP \over dz} = -\rho g_{\rm z} = -\rho \Omega_{\rm K}^2
z, }
\label{eq:stra}\\
\lefteqn{{d \varsigma \over dz} = 2 \rho,} \label{eq:strd} \\
\lefteqn{{d\ln T \over d  \ln P} = \nabla,} \label{eq:strb}\\
\lefteqn{{dF_{\rm z} \over dz } = {3 \over 2} \alpha_{\rm eff}
\Omega_{\rm K} P_{\rm g} } \label{eq:strc}
\end{eqnarray}
where $P=P_{g} + P_{\rm rad}$, $\rho$ and $T$ are the total (gas
plus radiation) pressure, density and temperature respectively,
$\varsigma$ is the surface column density between vertical
coordinates $-z$ and $+z$, $g_z = \Omega_{\rm K}^2 z$ the vertical
component of gravity, $\Omega_{\rm K}$ being the Keplerian angular
frequency, $F_z$ the vertical energy flux and $\nabla$ the
temperature gradient of the structure. This is generally radiative,
with $\nabla=\nabla_{\rm rad}$, given by:
\begin{equation}
\nabla_{\rm rad} = {\kappa P F_z \over 4 P_{\rm rad} c g_z},
\end{equation}
When the radiative gradient is superadiabatic, $\nabla$ is
convective ($\nabla=\nabla_{\rm conv}$). The convective gradient is
calculated in the mixing length approximation, in the same way as in
\citet{hmdl98}, with a mixing length taken as $H_{\rm ml} =
\alpha_{\rm ml} H_P$, where $H_P$ is the pressure scale height:
\begin{equation}
H_P = {P \over \rho g_z +(P \rho)^{1/2} \Omega_{\rm K} },
\end{equation}
which ensures that $H_P$ is smaller than the vertical scale height
of the disk. Here, we use $\alpha_{\rm ml}$ = 1.5.

Note that we have neglected the disk self gravity. This
approximation is valid as long as the ratio of self gravity to that
of the central object is small:
\begin{equation}
{g_{\rm s} \over g_{\rm c}} = {\Omega_{\rm K}^2 H_P \over 2 \pi G
\Sigma} < 1 \label{eq:selfgrav}
\end{equation}
If this not the case, the disk is gravitationally unstable, which,
as mentioned in the introduction, is likely to lead to fragmentation
if the cooling time is short enough, or may significantly change the
angular momentum transport by introducing non local terms \citep[see
e.g.][]{lp87,bp99}. In both cases, the thermal-viscous instability
can no longer apply (in the first case for obvious reasons, and in
the second one because non-local effects cannot be approximated by
viscosity, which is local); in our calculations we always make sure
that the condition (\ref{eq:selfgrav}) is fulfilled.

The parameter $\alpha_{\rm eff}$ is an effective viscosity, equal to
the standard viscosity coefficient $\alpha$ when the disk is in
thermal equilibrium, but which also accounts for the time-dependent
terms which are assumed to be also proportional to the pressure
\citep[see][for a detailed discussion]{hmdl98}.

The equation of state of matter is interpolated from the tables of
\citet{fgv77}; in the low temperature regime (below 2000 K), which
is not covered by these tables, Saha equations are solved
iteratively, as described by \citet{p69}. The Rosseland mean
opacities are taken from \citet{ct76} above 10,000 K, and from
\citet{a75} below \citep[more modern opacities introduce changes
that are not important in the present context, see][]{ldk}.

The boundary conditions are $\varsigma = 0$ and $F_z$ = 0 at the
disk midplane, and $\varsigma = \Sigma$ at the surface. The standard
photospheric condition $\kappa P_{\rm g} = 2/3 g_z$ has to be
slightly modified, as (1) radiation pressure can de dominant, and
(2) $g_z$ can vary in the photosphere. Integrating the vertical
hydrostatic equilibrium equation, and using the Eddington
approximation leading to $T^4(\tau) = 3/4 T^4_{\rm eff} ( 2/3 +
\tau)$ where $\tau$ is the optical depth, one obtains
\begin{equation}
\kappa \left(P_{\rm g} + {1 \over 2} P_{\rm rad}\right ) = {2 \over
3} g_z \left(1 + {1 \over \kappa \rho z} \right)
\end{equation}
The term $1/\kappa \rho z$ is of the order of the relative thickness
of the photosphere relative to the total disk thickness; it is
usually of little importance except when the disk luminosity is
close to its local Eddington limit, in which case the photosphere
can be quite extended.

\begin{figure*}
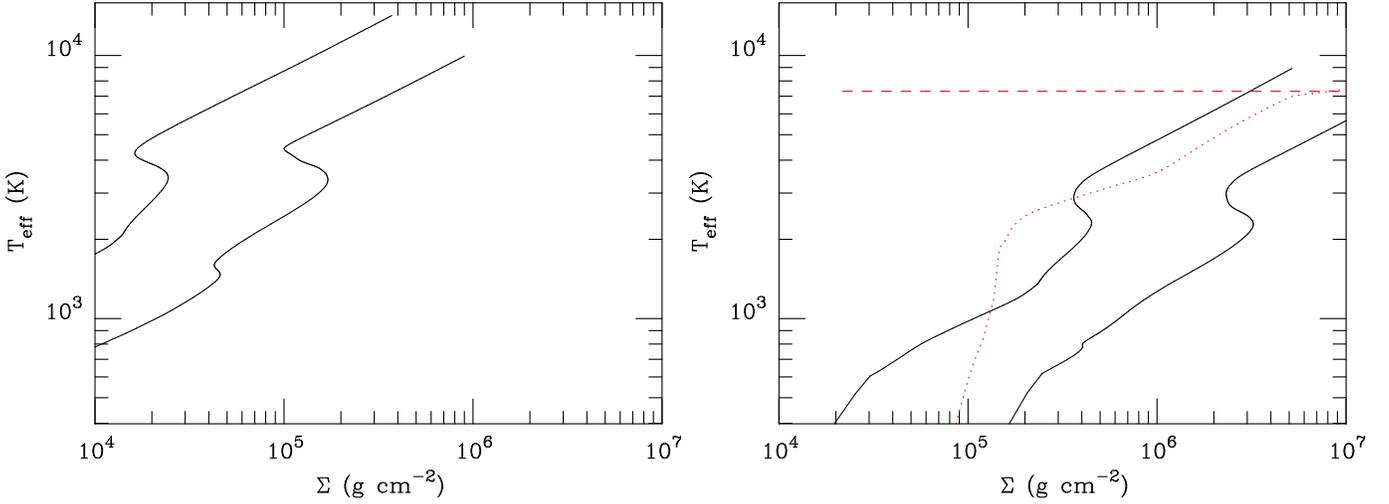

\resizebox{0.49\hsize}{!}{\includegraphics[angle=-90]{fig1a.eps}}
\resizebox{0.49\hsize}{!}{\includegraphics[angle=-90]{fig1b.eps}}
\caption{Examples of S curves in the $\Sigma$ -- $T_{\rm eff}$
plane, for $M=10^8 \rm M_\odot$ and $r = 10^{15}$ cm (left) and for
$r = 2 \times 10^{16}$ cm (right). In both cases, S curves obtained
for $\alpha$ = 0.1 and 0.01 are shown. The dotted curve correspond
to ratio of self to central gravity equals 1, and the dashed curve
to $h/r=0.1$. Only regions below the dashed curved and above the
dotted one are allowed. For $r = 10^{15}$ cm, these limits lay
outside the portion of the $\Sigma$ -- $T_{\rm eff}$ plane shown
here.}
 \label{fig:scurves}
\end{figure*}

The thermal equilibrium corresponds to $Q^+=Q^-$, where $Q^+$ and
$Q^-$ are the surface heating and cooling rates respectively (see
Eq. (\ref{eq:heat}) below). Figure \ref{fig:scurves} presents two
examples of thermal equilibrium curves in the $\Sigma$ -- $T_{\rm
eff}$ plane, showing the characteristic S shape. Also plotted are
the conditions $g_{\rm s} / g_{\rm c}=1$ and $h/r=0.1$. As can be
seen, self gravity becomes important at radii larger than $1 - 2
\times 10^{16}$ cm, in agreement with the findings of \citet{cr92}
and \citet{c92}. The condition that self-gravity be small can be
quite severe; we note for example that in several of the simulations
by \citet{jcs04} this condition is not fulfilled and the
corresponding results are therefore invalidated. The thin disk
approximation condition ($h/r \ll 1$) is usually less stringent; it
may however break for high accretion rates, in which case radiation
pressure gradient almost balances vertical gravity in a significant
fraction of the disk vertical extent.

The values $\Sigma_{\rm min}$ and $\Sigma_{\rm max}$ that are the
minimimun (resp. maximum) values of $\Sigma$ on the upper (resp.
lower) branches of the S curve can be fitted by:
\begin{equation}
\Sigma_{\rm min} = 2.90 10^3 \alpha^{-0.74}
               \left( {r \over 10^{15} \rm cm} \right)^{1.04}
               M_8^{-0.35}
               \rm g \; cm^{-2}
\end{equation}
and by:
\begin{equation}
\Sigma_{\rm max} = 3.85 10^3 \alpha^{-0.82}
               \left( {r \over 10^{15} \rm cm} \right)^{0.99}
               M_8^{-0.33}
               \rm g \; cm^{-2}
\end{equation}
where $M_8 = M/(10^8 \rm M_\odot)$, $M$ being the black hole mass;
the corresponding effective temperatures are $T_{\rm
eff}(\Sigma_{\rm min}) = 4300 (r/10^{15}\rm cm)^{-0.12}$ K and
$T_{\rm eff}(\Sigma_{\rm max}) = 3300 (r/10^{15}\rm cm)^{-0.12}$ K
respectively. Note that these are independent from $\alpha$, as
expected, and that their radial dependence is quite weak. As
compared to disks around stellar mass objects, the surface densities
are much higher and hence the effective temperature somewhat smaller
at the turning points of the $S$-curve (the upper stable solution
ends at 3000 -- 4000 K instead of 7000 -- 8000 K), even though the
corresponding mid-plane temperatures are quite similar.

\begin{figure}
\resizebox{\hsize}{!}{\includegraphics[angle=-90]{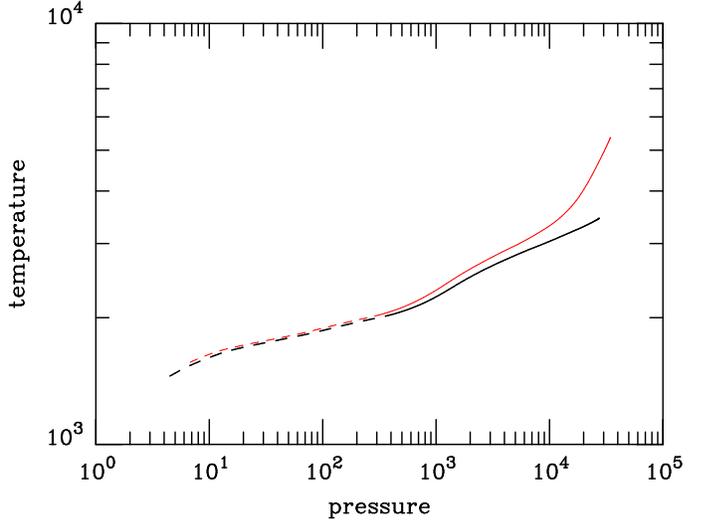}}
\caption{Vertical structure of the accretion disk, for $M=10^8 \rm M_\odot$,
$r=10^{15}$cm, $\Sigma=4.3\times 10^4$ g cm$^{-2}$, $\alpha = 0.01$ and
$T_{\rm c}=1400$ K (thick curve) and 1570 K (thin curve). The line
is dashed when energy transport is radiative. These two cases
correspond to the upper and lower stable part of the small wiggle
shown if Fig. \ref{fig:scurves}.}
 \label{fig:strvert}
\end{figure}

It should also be noted that, because we are restricted to a region
where self gravitation is small, the disk extension, as measured by
the ratio $r_{\rm in}/r_{\rm out}$ is not very large: for the case
of a 10$^8$ M$_\odot$ black hole, this is of order of 100, i.e.
comparable to disks in CVs, but much smaller than for LMXBs. It is
also worth noting that the disk thickness $H$ :
\begin{equation}
{H \over r} \simeq {c_{\rm s} \over v_{\rm k}},
\end{equation}
where $c_{\rm s}$ and $v_{\rm k}$ are the sound and Kepler
velocities, is small in regions where the thermal-viscous
instability can propagate. As compared to the CV or LMXB case,
$c_{\rm s}$ is unchanged because the central temperature at the
turning points of the S curve are similar, and of the order of
10$^6$ cm s$^{-1}$, but $v_{\rm k}$ is much larger since in the AGN
case we are restricted to regions close to the black hole as
mentioned above; here, we have $H/r \sim 10^{-3}$ or smaller. As the
width of the heating and cooling fronts are of order of a few times
$H$ \citep{mhs99}, this can be a source of numerical problems. In
particular, these fronts have been by far unresolved in all previous
studies, casting some doubt on their results.

\subsection{Critical points and the viscosity prescription}

It should be noted that for some choices of parameters the
equilibrium curves show two ``wiggles" (see e.g. the case $\alpha =
0.1$ at $r=10^{15}$ cm, Fig.\ref{fig:scurves}) on the lower branch.
This also happens in accretion disks around stellar mass black
holes, but, in contrast with \citet{jcs04}, we do not find that this
is always the case for AGNs. This discrepancy can be due to a
difference in the treatment of convection, or to different
opacities. The small wiggle at low temperature is not related to a
strong change in the opacities, but instead to a strong change in
the adiabatic gradient when molecular hydrogen becomes partially
dissociated, as is shown by Fig. \ref{fig:strvert}. Two vertical
structures, corresponding to the same $r$, $\Sigma$, and $\alpha$,
but two different effective temperatures on the upper and lower
stable branch of the small wiggle of Fig. \ref{fig:scurves}, differ
essentially by a much stronger temperature gradient in the disk
midplane: in one case, $\nabla_{\rm ad} \sim 0.10$, while in the
other $\nabla_{\rm ad} \sim 0.40$. This effect occurs only if
molecular hydrogen becomes partially dissociated in the convective
zone; since, for the corresponding temperatures and densities the
opacities are relatively low, there are cases where the transition
between molecular and atomic hydrogen occurs in a radiative zone, in
which case no wiggle is found.

In the standard dwarf-nova model, it is assumed that the
$\alpha$-parameter changes rapidly when the disk temperature reaches
the ionization instability; this is required for the amplitude of
the modeled outburst to be comparable to the observed one. It is
often stated that the physical reason for such a change is the
change in the ionization parameter of the gas, hence $\alpha$ is
assumed to remain constant when transiting this secondary wiggle.
This also seems to be a reasonable hypothesis in the AGN case, and
does not require the physics of accretion disk to be different in
different environments, even though the temperatures and densities
are similar. In the following contrary to \citet{jcs04}, we assume
therefore that the critical $\Sigma_{\max}$ of the cold stable
branch corresponds to the ionization instability and that the lower
wiggle is not associated with a change in $\alpha$. This point is of
importance, since as shown by \citet{h02}, the shape of the
resulting S curve and hence the outcome of the model is by far
dominated by the change in the viscosity parameter $\alpha$.

Finally, one should remark that whatever arguments are used to
justify the change in $\alpha$, i.e. the use of an $\alpha_{\rm
cold}$ and an $\alpha_{\rm hot} \approx (4 - 10) \alpha_{\rm cold}$
the real reason is the necessity to produce the required outburst
amplitude. It has been argued \citep{gm98} that the difference
between viscosities in the high and low (quiescent) states of
dwarf-nova disks is due to the ``decay" of the MRI mechanism  that
is supposed to be the source of turbulence in accretion disks
\citep{bh91}. In the environment of AGN disks, the MRI is supposed
to be operating also in cold disks, which was used to argue that in
this case $\alpha_{\rm hot} \approx \alpha_{\rm cold}$ \citep{mq01}.

However, as noted by Steven Balbus (private communication), because
of the fact that numerical simulations treat the turbulent dynamics
of disks at a level far beyond anything that can be approached with
strictly analytic techniques, there has been a tendency to grant
simulations a level of certainty that they do not yet merit. A
careful treatment of realistic energetics still remains beyond the
capabilities of current codes, and even simple polytropic shearing
box calculations need to be run at much higher resolutions and for
much longer times than were once thought necessary.

Therefore the values of critical Reynolds numbers deduced from
numerical simulations only  \citep{gm98,mq01} are highly uncertain
and we opted for using the standard dwarf-nova DIM also in AGNs.

\section{Disk evolution}

\subsection{Basic equations}

The standard equations for mass and angular momentum conservation in
a geometrically thin accretion disk can be written as:
\begin{equation}
{\partial \Sigma \over \partial t} = - {1 \over r} {\partial \over
\partial r} (r \Sigma v_{\rm r})
\label{eq:consm}
\end{equation}
and
\begin{equation}
j{\partial \Sigma \over \partial t} = - {1 \over r} {\partial
\over \partial r} (r \Sigma j v_{\rm r}) + {1 \over r} {\partial
\over \partial r} \left(- {3 \over 2} r^2 \Sigma \nu \Omega_{\rm
K} \right)
\label{eq:consj}
\end{equation}
where $v_{\rm r}$ the radial velocity in the disk, $j =
(GM_1r)^{1/2}$ is the specific angular momentum of material at
radius $r$ in the disk, $\Omega_K=(GM_1/r^3)^{1/2}$ is the Keplerian
angular velocity

The energy conservation equation is taken as \citep[see][for
details]{c93,hmdl98}:
\begin{equation}
{\partial T_{\rm c} \over \partial t} = { 2 (Q^ + -Q^- + J) \over
C_P \Sigma}
 - {P_{\rm c}
\over \rho_{\rm c} C_P} {1 \over r} {\partial (r v_{\rm r}) \over
\partial r} - v_{\rm r} {\partial T_{\rm c} \over \partial r},
\label{eq:heat}
\end{equation}
where $P_{\rm c}$ and $\rho_{\rm c}$ are the midplane pressure and
density, and $Q^+$ and $Q^-$ are the surface heating and cooling
rates respectively. They are usually taken as $Q^+=(9/8) \nu \Sigma
\Omega_{\rm K}^2$ and $Q^- = \sigma T_{\rm eff}^4$, $T_{\rm eff}$
being the effective temperature. The term $J$ accounts for the
radial energy flux carried by viscous processes:
\begin{equation}
J = 1/r \partial / \partial r (r F_{\rm e}).
\end{equation}
where $F_{\rm e}$ is the flux carried in eddies with
characteristic velocity $v_{\rm e}$ and size $l_{\rm e}$, is:
\begin{equation}
F_{\rm e} = C_P \Sigma v_{\rm e} {\partial T_{\rm c} \over
\partial r} l_{\rm e} = {3 \over 2} \nu C_P \Sigma {\partial
T_{\rm c} \over \partial r}, \label{eq:fturb}
\end{equation}
These are identical to the equations of a disk in a binary system,
except that there are no tidal torques and no tidal dissipation.

The inner boundary condition is also unchanged from the binary
case:
\begin{equation}
\nu \Sigma = 0 \;\; {\rm at} \; r=r_{\rm in}
\end{equation}
where $r_{\rm in}$ is the radius of the inner edge of the disk, and
can be larger than the radius of the innermost stable orbit if the
disk is truncated by the formation of an ADAF, in which case $r_{\rm
in}$ is a given function of the mass accretion rate \cite[see
e.g.][]{hlmn97}. As we use $\ln (\Sigma)$ as a variable, the $\Sigma
= 0$ boundary condition is not applicable. Instead, we take:
\begin{equation}
\nu \Sigma = 1.1 \Sigma_{\rm min} \;\; {\rm at} \; r=r_{\rm in}
\end{equation}
so that this allows both for the disk to be in the hot or cold state
(see below for a more detailed discussion on the effect of using
this boundary condition).

The outer boundary condition is more problematic, as the disk
extends to large distances where all the usual approximations are
invalid (thin disk, neglect of self gravity, ...). We instead assume
that at some distance $r_{\rm out} \sim 10^{16}$ cm for $M = 10^8$
M$_\odot$, the mass transfer rate is given and constant. This
approximation is valid provided that the heating front does not
reach this outer radius.

The heat equation (\ref{eq:heat}) requires two additional boundary
conditions; as discussed in \citet{hmdl98}, these are of little
importance, and we take $J=0$ at $r=r_{\rm in}$ and $r=r_{\rm out}$.

\begin{figure}
 \resizebox{\hsize}{!}{\includegraphics{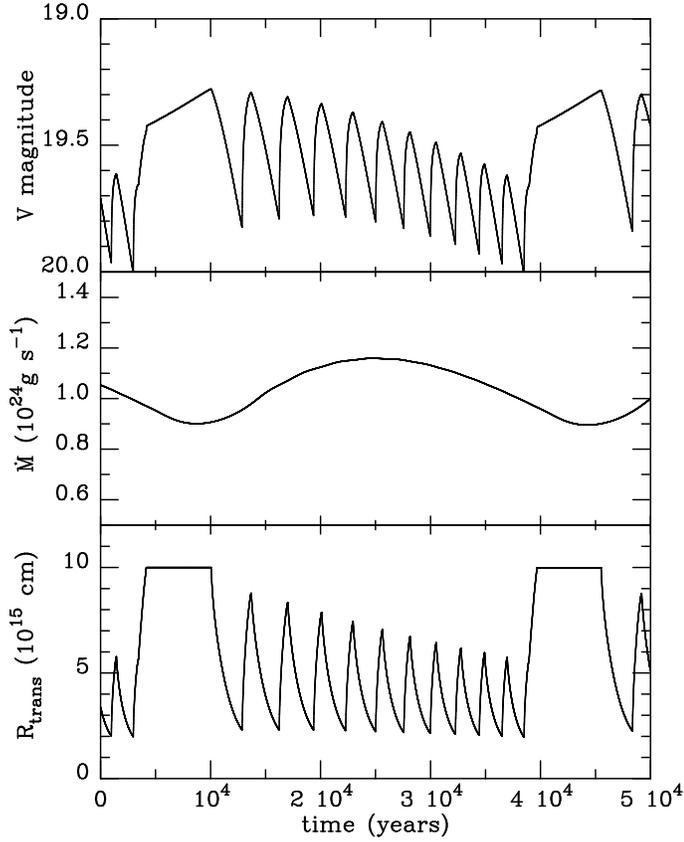}}
 \caption{Time evolution of an accretion disk with the following
 parameters: black hole mass: 10$^8$ M$_\odot$, inner and outer
 radius: 10$^{14}$ and 10$^{16}$ cm respectively, and mean mass
 transfer rate: 10$^{24}$ g s$^{-1}$. Top panel: visual magnitude,
 intermediate panel: accretion rate onto the black hole, lower
 panel: radius at which the transition between the hot and cold
 regimes takes place.}
 \label{fig:std}
\end{figure}

\subsection{Results}

\begin{figure}
 \resizebox{\hsize}{!}{\includegraphics{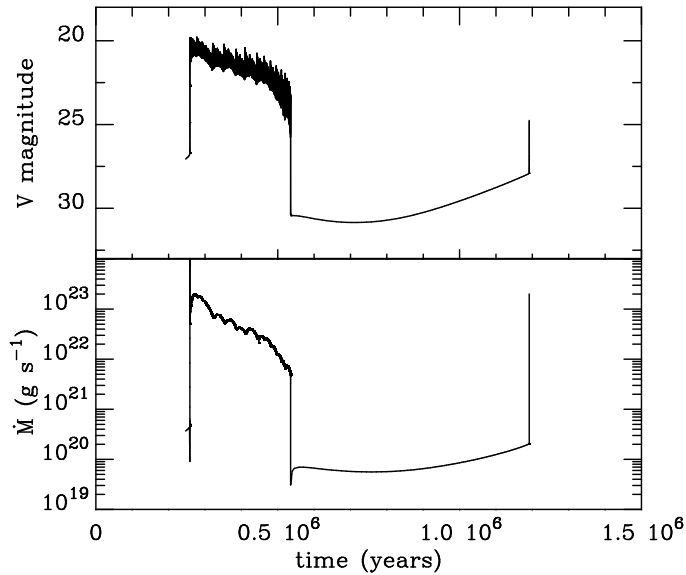}}
 \caption{Time evolution of an accretion disk with the following
 parameters: black hole mass: 10$^8$ M$_\odot$, inner and outer
 radius: 10$^{14}$ and 10$^{16}$ cm respectively, and mean mass
 transfer rate: 2 10$^{22}$ g s$^{-1}$. Top panel: visual magnitude,
 lower panel: accretion rate onto the black hole.}
 \label{fig:lowmdot}
\end{figure}

\begin{figure}
 \resizebox{\hsize}{!}{\includegraphics{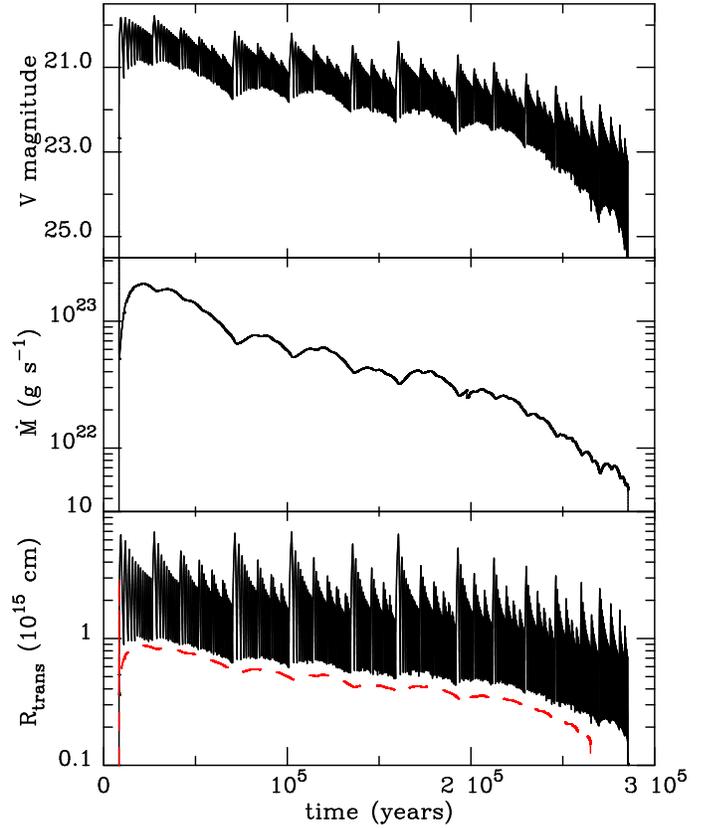}}
 \caption{Details of the outburst shown in Fig. \ref{fig:lowmdot}.
 Top panel: visual magnitude,
 intermediate panel: accretion rate onto the black hole, lower
 panel: radius at which the transition between the hot and cold
 regimes takes place.The red-dashed line is the semi-analytic value
 of the minimum transition radius given by Eq. \ref{eq:rmin}}
 \label{fig:lowmdot_details}
\end{figure}

Figure \ref{fig:std} shows an example of the evolution of the
accretion disk. We have considered here a 10$^8$ M$_\odot$ black
hole accreting at 10$^{24}$ g~s$^{-1}$, about one hundredth of the
Eddington limit:
\begin{equation}
\dot{M}_{Edd} = {L_{\rm Edd} \over c^2 \eta} = {4 \pi G M m_{\rm p}
\over \sigma_{\rm th} c \eta} \simeq 1.4 \times 10^{26} M_8 \rm g \;
s^{-1}
\end{equation}
where $\eta \sim 0.1$ is the efficiency of accretion, and
$\sigma_{\rm Th}$ is the Thompson cross section. The mass is the
same as in \citet{jcs04}, but the mass transfer rate is lower by a
factor $\sim$ 10 in order to avoid (or try to avoid) the instability
to occur in the self gravitating part of the accretion disk. We have
assumed a varying $\alpha$ between $\alpha_{\rm h} = 0.2$ and
$\alpha_{\rm c} = 0.04$. We have also taken the outer disk radius to
be $10^{16}$ cm, as the disk becomes self gravitating at larger
distances. As can be seen, the disk can never be brought completely
to the cold regime; as a consequence, relatively low amplitude
oscillations are seen in the visual magnitude and in the accretion
rate onto the black hole. This situation is reminiscent of what
happens in the case of soft X-ray transients when no disk truncation
or irradiation is assumed \citep{mhl00,dhl01}, or in the case of
symbiotic stars \citep{d86}: a cooling front is reflected at some
radius much larger than the disk inner radius and as a result a
heating front starts propagating outwards but it cannot quite reach
the outer disk edge and a cooling front forms again. Such
reflections occur when the surface density $\Sigma$ behind the
cooling front reaches $ \Sigma_{\rm max}$ which triggers a new
instability. The resulting heating front propagates outwards until
the post-front density reaches $\Sigma_{\rm min}$. Then a new
cooling front starts going down the disk. There is however a
significant difference in that the short time oscillations do not
result in oscillations of the mass accretion rate on the same time
scale. $\dot{M}$ fluctuates only on the longer time scale of the
front oscillation pattern. The basic reason for this is that the
front propagates at approximately $\alpha$ times the sound speed,
i.e. on a time scale
 \begin{equation}
 t_{\rm front} = {r \over \alpha c_{\rm s}} = {r \over h} t_{\rm th},
 \end{equation}
where $t_{\rm th}$ is the thermal time scale. $t_{\rm front}$ is
shorter than the viscous time $t_{\rm visc} = (r/h)^2 t_{\rm th}$ by
a factor $r/h$, i.e. by several orders of magnitude. The cooling
front therefore propagates so rapidly that the surface density at
smaller radii does not change; to a first approximation, it cannot
propagate in regions where $\Sigma > \Sigma_{\rm max} (\alpha_{\rm
c})$. In the CV case, $t_{\rm front}$ is shorter than $t_{\rm
visc}$, but not by such a large amount and strong gradients in the
disk make the effective viscous time comparable to the front
propagation time.

It must also be noted that the front occasionally reaches the outer
disk edge; then the outer boundary condition which imposes in
particular that there is no outward mass flow is not valid, so that
the correct sequence is probably different; the back and forth
propagation of heating fronts on a short time scales is however a
firm prediction of the model.

\begin{figure}
 \resizebox{\hsize}{!}{\includegraphics[angle=-90]{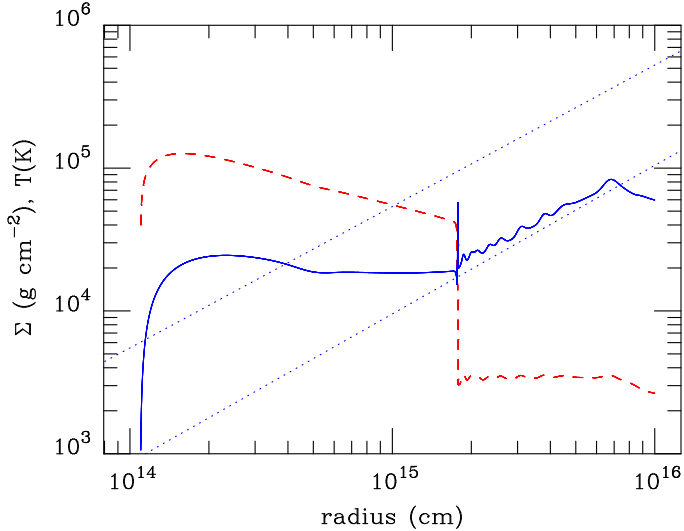}}
 \caption{Radial structure of the disk. The blue solid line represents
 the surface density, the red dashed one the central temperature.
 The dotted lines are the critical $\Sigma_{\rm min}$ and
 $\Sigma_{\rm max}$. See text for details.}
\label{fig:strucrad}
\end{figure}

For lower mass transfer rates, the outer part of the disk can remain
on the cold, stable branch, in which case the front propagation is
restricted to the innermost parts of the disk. Figures
\ref{fig:lowmdot} and \ref{fig:lowmdot_details} show the evolution
of disk with the same parameters as in Fig. \ref{fig:std}, but with
a mass transfer rate of 2 10$^{22}$ g s$^{-1}$, 5 10$^{-3}$ times
the Eddington limit. As can be seen in Fig.
\ref{fig:lowmdot_details}, heating and cooling fronts propagate in a
restricted fraction of the accretion disk. They do not reach radii
larger than that at which the disk can sit on the stable cool
branch, given the externally imposed mass transfer rate. They also
do not reach the innermost regions where the surface density always
remain high enough for the disk to be stable on the hot branch,
except when entering a quiescence periods, which happens when the
disk finally empties on a much longer viscous time. The active phase
lasts for 3 10$^5$ yr in the case presented here, with more than 400
consecutive oscillations. Note that these are not random, but show
relatively regular sequences of decreasing oscillations that are
interrupted by an oscillation with a larger amplitude, clearly
visible in Fig. \ref{fig:lowmdot_details}. Note also some sort of
hierarchical structure of these oscillations. Figure
\ref{fig:strucrad} shows the radial structure of the disk during the
oscillating phase \citep[compare wih Fig. 2 of][]{dhl01}. The
semi-stable inner and outer regions are clearly visible; the
unstable zone is in fact divided in two regions: an inner unstable
one, and an outer marginally stable one, where $\Sigma \approx
\Sigma_{\rm min}$, resulting from the successive passage of heating
fronts that die at radii decreasing with time; a leftover of the
death of these fronts is the little wiggle in $\Sigma$ that gets
smoothed with time as a result of diffusion, or when a heat front is
able to reach this region. Note also the spike in the unstable
region, that carries a small amount of mass that will cause the
small wiggles in the marginally stable region.

\subsection{Minimum radius reached by cooling fronts}

The minimum radius reached by the cooling front can be determined by
noting that the front propagates down to a point where $\Sigma =
\Sigma_{\rm max}(\alpha_{\rm c})$, and that the innermost parts of
the disk are in quasi viscous equilibrium, which means that the
surface density is determined by the accretion rate, almost constant
in this hot inner region. This is equivalent to stating that, at the
reflection point, the dissipation rate $Q^+$ is:
\begin{equation}
Q^+ = {3 G M \dot{M} \over 8 \pi r^3} f
      = \sigma T_{\rm eff}^4(\Sigma_{\rm max}, {\rm hot})
\end{equation}
where $f=1-(r/r_{\rm in})^{-1/2}$. Note that $T_{\rm eff}$ is
calculated on the hot branch, and is {\em not} given by the analytic
fits obtained in Section 2. An examination of Fig. \ref{fig:scurves}
shows that $T_{\rm eff}(\Sigma_{\rm max}, {\rm hot})$ is about 3.2
times that of the turning point on the cool branch, $T_{\rm
eff}(\Sigma_{\rm max}, {\rm cold})$ for $\alpha_{\rm c}/\alpha_{\rm
h} = 0.1$. As $Q^+$ is proportional to $\nu \Sigma$, and hence
proportional to $\alpha$, one can guess that:
\begin{equation}
T_{\rm eff}^4(\Sigma_{\rm max}, {\rm hot}) = 3.2^4 {\alpha_{\rm h}
\over 10 \alpha_{\rm c}} T_{\rm eff}^4(\Sigma_{\rm max}, {\rm cold})
\end{equation}
which is also a very good approximation even for $\alpha_{\rm
c}/\alpha_{\rm h} = 1$, as can be seen from Fig. \ref{fig:scurves}.
Now, from the fits of $T_{\rm eff}^4(\Sigma_{\rm max}, {\rm hot})$
we have:
\begin{equation}
r= 1.7 \times 10^{15} \left( {\dot{M} \over 10^{23} \rm g \; s^{-1}}
M_8 {\alpha_{\rm c} \over \alpha_{\rm h}} f \right)^{0.4}
\label{eq:rmin}
\end{equation}

\begin{figure}
 \resizebox{\hsize}{!}{\includegraphics{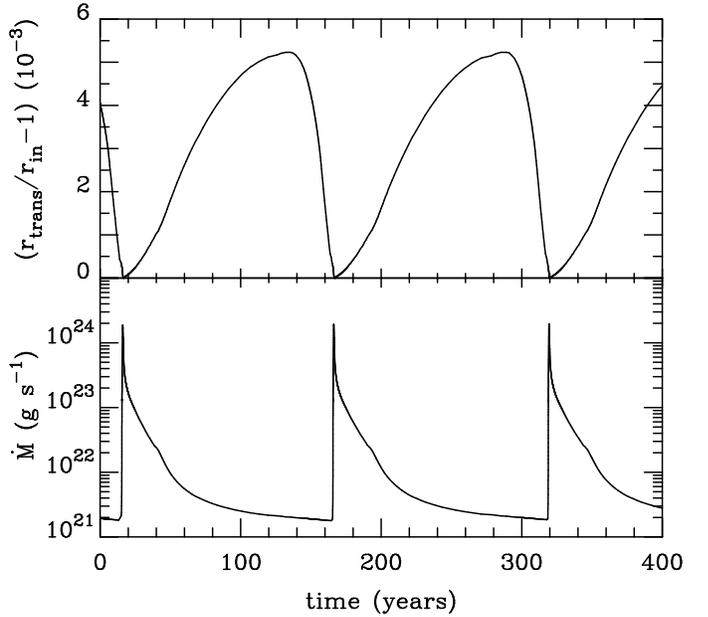}}
 \caption{Instability of the inner disk edge when the boundary condition
 $\Sigma=0$ is used. The top panel shows the position of the transition
 radius as a function of time, the lower panel the mass accretion rate
 onto the black hole. The parameters are those of Fig. \ref{fig:lowmdot}.
 These large amplitude are unphysical, since mass flows into the black hole
 from a region much larger than the width of the zone in which these fluctuations occur.}
\label{fig:sigma0}
\end{figure}

Figure \ref{fig:strucrad} shows that there are two points where
$\Sigma$ crosses the $\Sigma_{\rm max}$ line, and Eq. \ref{eq:rmin}
has indeed two solutions, one for which $f$ is small, and another
one in which $f \simeq 1$. The first one corresponds to the
transition between the very inner disk, where $\Sigma$ is
vanishingly small because of the boundary condition and therefore
the cool branch solution applies, and nearby regions where $\Sigma$
is large enough for the hot solution to apply. This will be
discussed in the next section. The second one corresponds to the
radius at which the cooling front is reflected into a heating front.

\begin{figure}
 \resizebox{\hsize}{!}{\includegraphics{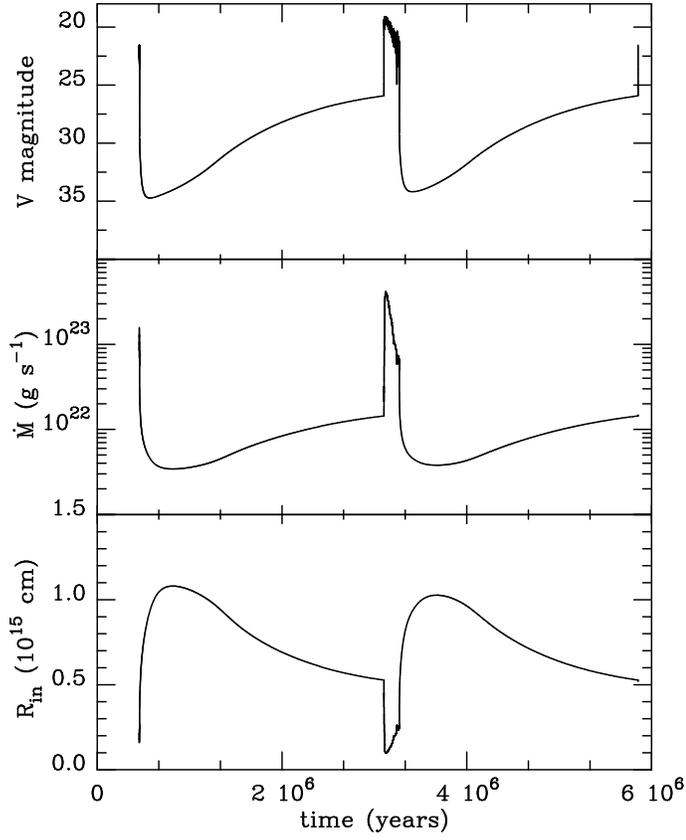}}
 \caption{Long term evolution on an AGN accretion disk when $r_{\rm in}$
 can vary as a result of
 e.g. evaporation or the formation of an ADAF.
 The system alternates between active
 phase in which heating and cooling fronts propagate back and forth
 in the disk and quiescent phases lasting about one million years.
 The top panel: visual magnitude, intermediate panel: accretion rate
onto the black hole; lower panel: inner disk radius.}
 \label{fig:evap2}
\end{figure}

\begin{figure}
 \resizebox{\hsize}{!}{\includegraphics{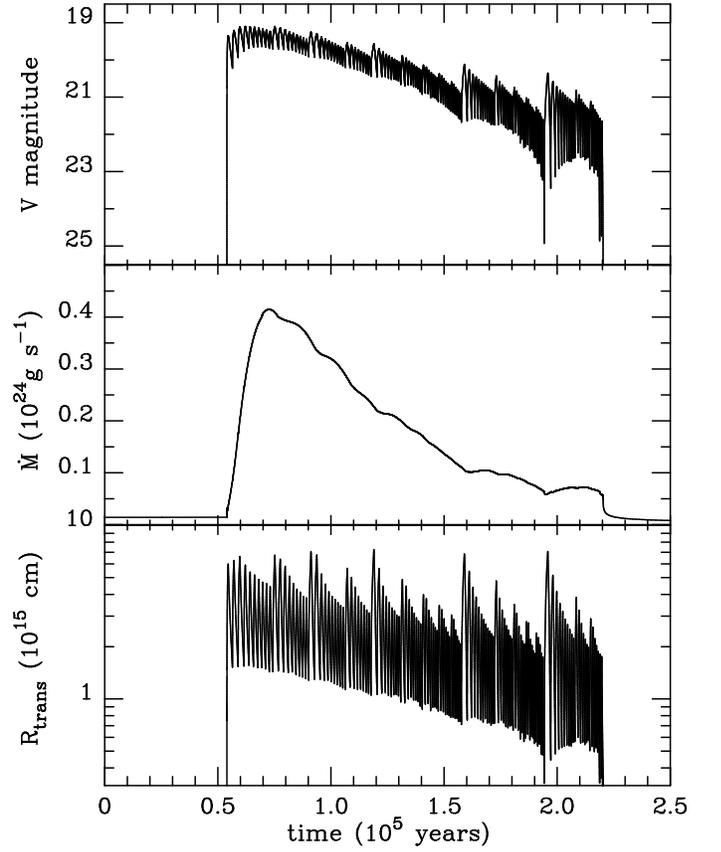}}
 \caption{The outburst shown in Fig. \ref{fig:evap2}. The two upper panels
 are the same as in Fig. \ref{fig:evap2}, the lower panels shows the
 transition radius between the inner hot disk and the cooler outer
 parts}
 \label{fig:evap}
\end{figure}

From Eq. \ref{eq:rmin}, it appears that for low enough $\dot{M}$ or
large enough $r_{\rm in}$, the cooling front can reach the inner
radius, in which case the system will enter a quiescence phase. More
precisely, this happens when Eq. \ref{eq:rmin} has no solution.
Simple algebra shows that the critical $\dot{M}$ is:
\begin{equation}
\dot{M} = 5 \times 10^{20} \left( {r_{\rm in} \over 10^{14} \rm
cm}\right)^{2.5} {\alpha_{\rm h} \over \alpha_{\rm c}} M_8^{-1} \;
\rm g \; s^{-1} \label{eq:mmin}
\end{equation}
and the corresponding critical radius is $r = 1.44 R_{\rm in}$, at
which $f=1/6$. It is interesting to rescale this relation as:
\begin{equation}
{\dot{M} \over \dot{M}_{\rm Edd}} = 2.7 10^{-6} {\alpha_{\rm h}
\over \alpha_{\rm c}} M_8^{0.5} \left( {r_{\rm in}/3 r_{\rm s}}
\right)^{2.5} \label{eq:mmin1}
\end{equation}
which shows that, if the disk is not truncated, low states will be
found only for low mass transfer rates, whatever the black hole
mass.

The critical rate given by Eq. \ref{eq:mmin1} refers to the
accretion rate onto the black hole, and not to the mass transfer
rate. For high mass transfers rates, both are almost equal, as seen
above (see e.g. Fig. \ref{fig:std}); for lower values of
$\dot{M}_{\rm transf}$, they may differ by up to one order of
magnitude, as shown for example in Fig. \ref{fig:lowmdot} where the
accretion rate at maximum is 10 times larger than the mass supply
rate to the disk. In this case, $\dot{M}$ ultimately falls below the
critical value given by Eq. \ref{eq:mmin1}, and the disk enters a
quiescent state. The duration of this state is short, however, of
order of the duration of the outburst state, as the average mass
accretion rate during the active state is $\sim 5.7 \; 10^{22}$g
s$^{-1}$, i.e. not very different from the steady mass transfer rate
(note that the duty cycle expected for an outburst with an average
accretion rate 2.85 times larger than the transfer rate is 0.35,
very close to the value given by the simulation (0.30), showing that
the disk is almost relaxed). We therefore expect that outbursts
exist only for low mass transfer rates, that these outbursts are
weak -- never reaching anything close to the Eddington limit -- and
that the duty cycle cannot be large.

\subsection{Innermost disk instability}

The very inner parts of the disk, where the density is very low
because of the inner boundary condition should therefore be on the
cold branch; the transition between this cold region and more
distant, hotter regions should also be unstable. Indeed, when one
assumes that the inner boundary condition is not $\Sigma = 1.1
\Sigma_{\rm min}$ at $r = r_{\rm in}$, but is smaller than
$\Sigma_{\rm min}$, oscillations are found. For the sake of
completeness, we show in Fig. \ref{fig:sigma0} the effect of such
oscillations in such a case. Cooling/heating fronts propagate in a
very restricted region, whose radial extent is larger than the
vertical scale height so that the thin disk approximation is still
valid, but much presumably smaller than the zone from which matter
flows into the black hole. Note also that this region is so small
that the total disk luminosity remains constant. These oscillations
are possible only when the width of heating/cooling fronts is
smaller than the width of the region over which $\Sigma$ catches the
boundary condition $\Sigma=0$, which is of the order of one to a few
percents of $r_{\rm in}$ (see Fig. \ref{fig:strucrad}), otherwise
fronts would simply not exist. This is quite possible in the AGN
case, because the fronts are so narrow and contrasts with the CV or
LMXB case where the reverse is true and the condition $\Sigma = 0$
does not have such an effect. There oscillations are most probably
not physical, because one assumes that (i) there is absolutely no
torque at the inner disk edge, and (ii) that matter is lost from the
disk only at $r=r_{\rm in}$; it is very likely that the mechanisms
leading to accretion at the inner disk edge (e.g. evaporation, ...)
will smooth oscillations there. In order to avoid these, and in
order to ease the numerical computations, we have assumed that
$\Sigma$ is not vanishingly small at the inner disk edge, but that
instead it is very slightly larger than $\Sigma_{\rm min}$.

\subsection{Disk truncation}

Disk truncation could be a solution to the absence of large
outbursts; this was found to ba an essential ingredient of the soft
X-ray transient model \citep[see e.g.][]{mhl00,dhl01}. Truncation
can be the result of the formation of an advection-dominated
accretion flow (ADAF), or of one of its variants \citep[see
e.g.][for reviews of the ADAF]{nm08,kfm98}; the important feature
being that the flows becomes hot, geometrically thick, and optically
thin close to the black hole. In order for the outburst cycle to be
modified, one needs the disk not to extend down to the innermost
stable orbit, but instead be truncated at a radius comparable to the
minimum radius reached by the cooling front. The inner disk radius
will then depend on the mass accretion rate onto the black hole.
Many prescription can be derived; what really matters is whether Eq.
(\ref{eq:rmin}) can be satisfied or not, the details of the
variations of $r_{\rm in}$ as a function of $\dot{M}$ being of no
importance.

Figures \ref{fig:evap2} and \ref{fig:evap} show an example in which
$r_{\rm in}$ = 2 10$^{14}$ $(\dot{M}/10^{23} \rm \; g \;
s^{-1})^{1/2}$. As can be seen, quiescent states are found, as well
as active state which are not very bright though -- only brighter
than the active states for non truncated disks by a factor $\sim$ 2,
resulting in duty cycles which also differ by factors $\sim$ 2.

\subsection{Disk irradiation}

\begin{figure}
 \resizebox{\hsize}{!}{\includegraphics{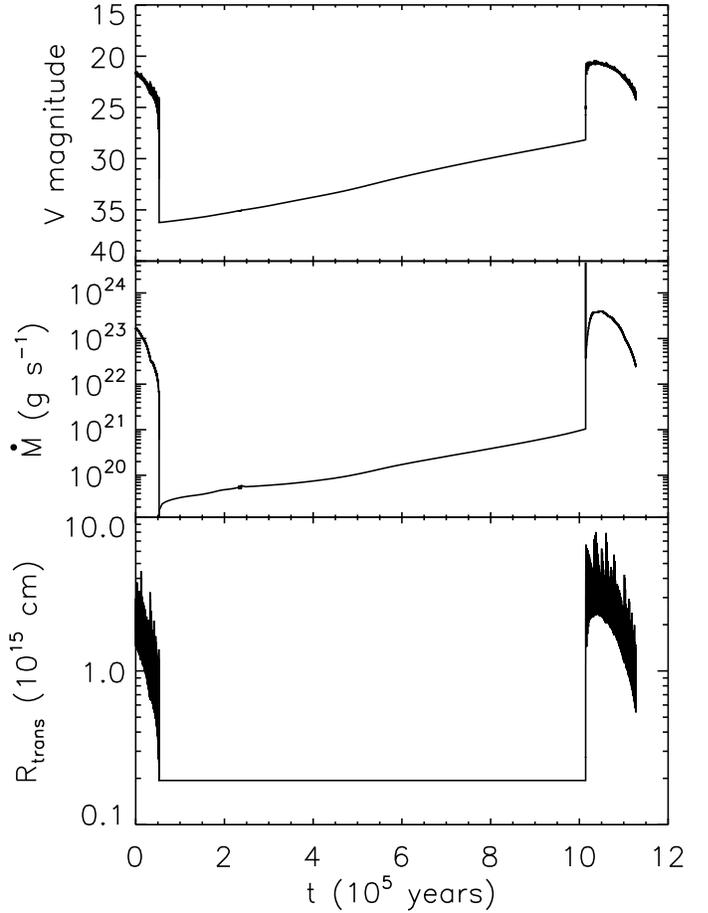}}
 \caption{Time evolution of an irradiated disk, with parameters
 identical to those of Fig. \ref{fig:lowmdot}; here $C=0.01$. Top
 panel: V magnitude; intermediate panel: accretion rate onto the
 black hole; lower panel: transition radius between the hot, inner
 disk and the cool outer disk.}
 \label{fig:evol_irr}
\end{figure}

\begin{figure}
\resizebox{\hsize}{!}{\includegraphics[angle=90]{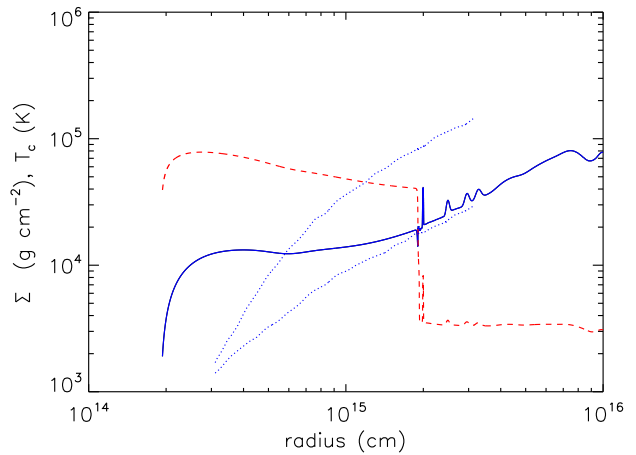}}
\caption{Example of the disk radial structure in the irradiated
case. The blue solid line is the surface density, the red dashed
line is the central temperature. The blue dotted lines represent
$\Sigma_{\rm min}$ and $\Sigma_{\rm max}$, and the effects of
irradiation are clearly visible ar radii $\lta$ 10$^{15}$cm; at
radii smaller than 3 10$^{14}$cm, the S shape of the cooling curve
vanishes and the disk is stable}.\label{fig:str_irr}
\end{figure}

Disk irradiation plays an essential role in soft X-ray transients
\citep[see e.g.][and references therein]{vp96,dhl01}, and sometimes
in CVs, see e.g. \citep{hld99}; it could also play an important role
in the AGN context. We follow here the same procedure as in the case
of irradiated disks in SXTs \citep{dhl01}. We assume that the
irradiation flux $F_{\rm irr}$ onto the disk is given by:
\begin{equation}
\label{eq:tirr_agn}
 F_{\rm irr} = \sigma T_{\rm irr}^4 = C \frac{\dot{M} c^2}{4\pi R^2}
\end{equation}
where $C$ is a constant; in the case of SXTs, $C = 5 \times 10^{-4}$
has been adopted by \cite{dhl01} (note that we include here in $C$
the efficiency of mass to energy conversion). The ratio of the
irradiating flux to the viscous flux is then given by
\begin{equation}
 {F_{\rm irr} \over F_{\rm visc}} =  {4 \over 3} C {\dot{M}_{\rm acc}
 \over \dot{M}} {r \over r_{\rm s}}
 \label{eq:irr_effect}
\end{equation}
which clearly shows, because we are interested in regions much
closer to the black hole than in the case of SXTs, that $C$ must be
large if irradiation is to have any effect at all: in order to
affect the central temperature of the disk, $F_{\rm irr} / F_{\rm
visc}$ must exceed the optical thickness of the disk (in the
radiative case). This is in principle possible because the X-ray
emitting region could have a complex geometry, as e.g. a corona
above a cooler disk, in which case the irradiation flux can be
large; it is however very unlikely that it could exceed 0.01, since
the fraction of X-rays absorbed below the photosphere is at most
10\%. In the following, we consider the case $C=0.01$, which
corresponds to a maximally irradiated disk. One should also note
that in this case, the X-ray luminosity would be linked in a complex
way to the local properties of the accretion flow, and would not be
proportional to the accretion rate onto the black hole $\dot{M}_{\rm
acc}$ only. Note also that Eq. (\ref{eq:irr_effect}) assumes steady
state, and neglects the $r_{\rm in}/r$ terms in the energy
dissipation equation, which can be significant close to the disk
inner edge.

We have then calculated a grid of vertical disk structures in order
to determine the effective temperature as a function of $\Sigma$,
$T_{\rm c}$, $T_{\rm irr}$, as described in Section 2, with a
modified boundary condition at the disk surface
\begin{equation}
F_z = \sigma (T^4 -T_{\rm irr}^4)
\end{equation}
As in the case of SXTs, the effect of irradiation is a stabilization
of the disk when the irradiation temperature is large enough,
typically larger than $\sim$ 10$^{4}$K.

A reasonably good fit to the effective temperature at $\Sigma_{\rm
max}$ is given by:
\begin{eqnarray}
\label{eq:teff_coude_irr}
 T_{\rm eff}(\Sigma_\mathrm{max}) & = & 730 \;
 \Big( \frac{r}{10^{15}\mathrm{cm}}\Big)^{0.33} M_8^{-0.11} \Big(
 \frac{M_{\rm acc}}{10^{25}\mathrm{g\ s}^{-1}}\Big)^{-0.29}  \nonumber \\
 & & \times \Big \{ 1 - \Big(\frac{7\times10^{19} \mathrm{g\ s}^{-1}}{M_{\rm
 acc}}\Big)^{1/4} \Big \}^{1/4}\mathrm{\ K}
\end{eqnarray}
in a situation where the disk is in viscous equilibrium and hence
$T_{\rm irr}$ is directly given by $\dot{M}$, valid for irradiation
temperatures larger than about 4000K, but less than 10$^4$K for
which the disk becomes stable.

Figure \ref{fig:evol_irr} shows the time evolution of an irradiated
disk with the same parameters as in Fig. \ref{fig:lowmdot}, apart
from the irradiation factor $C$ set to 0.01. As can be seen, even in
this maximally irradiated disk, the time evolution is not very
different from that of unirradiated disk. There is still a
succession of rapid oscillations of the luminosity, with a
heating/cooling front propagating back and forth; the main
difference is here that the disk enters into a quiescent phase more
rapidly than in the unirradiated case, but one should note that the
initial structure was not exactly the same in both cases, and that
because of the huge computing time required to follow the disk
oscillations, a relaxed state can not be attained in practice.
However, the radial disk structure obtained at the end of active
phases in the irradiated and unirradiated case, with an outer disk
on the cold stable branch, and most of the disk having $\Sigma =
\Sigma_{\rm min}$; this similarity, and the fact that the
unirradiated disk was almost relaxed makes us confident that here
also the disk is close to relaxation.

Figure \ref{fig:str_irr} shows the radial structure of the disk at
time $t= 3.5 \; 10^4$yr, when 3/4 of the first outburst have
elapsed. It clearly shows the impact of irradiation on the innermost
part of the disk, which is due both to a large irradiation
temperature and to the decreasing viscous dissipation close to the
inner disk edge (the $f$ factor).

The conclusion that quiescent states are possible only for low mass
transfer rates is very general; an analysis similar to that
described above in the non irradiated case leads to the conclusion
that the disk can enter into quiescence only if the mass accretion
rate is less than
\begin{equation}
\label{eq:mcrit_irr} \dot M_\mathrm{acc,crit} = 4\times 10^{20} \Big
( \frac{r_\mathrm{in}}{10^{14}\mathrm{cm}}\Big)^{2} M_8^{-2/3} \Big
( \frac{\alpha_h}{\alpha_c}\Big )^{0.46} \mathrm{\ g\ s}^{-1}
\end{equation}
accurate to within a factor 2 when compared to the results of
the numerical simulations. Although the dependence on $M_8$,
$r_{\rm in}$, and $\alpha$ are different from the unirradiated
case, the numerical value of this critical rate is not changed
to the point where one could obtain large amplitude outbursts,
during which the Eddington limit would be attained or
approached.

\section{Conclusion}

We have shown that the accretion disks in active galactic nuclei can
indeed be subject to the same thermal-viscous instability as in
dwarf novae and soft X-ray transients, but the outcome of this
instability is very different. This contrasts with previous findings
that large amplitude outbursts reaching the Eddington limit were
possible, and the reason for this discrepancy is the insufficient
spatial resolution of the numerical codes that have been used to
model the disk. In AGNs, the disk opening angle is much less than in
DNs or SXTs, because the Keplerian velocity is not small compared to
the speed of light, whereas the sound speed is, by construction, the
same in both cases. This results in very thin transition fronts
which are quite difficult to follow numerically.

We do however predict time variations of the AGN luminosity by a few
magnitudes on time scales ranging from a few thousand (the
propagation time of a thermal front in the disk) to a few million
years (the typical quiescent/outburst time), and these oscillations
are enhanced by a possible truncation of the innermost parts of the
disk. However, because of their too small amplitudes and duty
cycles, these variations cannot explain the statistical properties
of quasar and AGN luminosity distribution.

One should also stress that, for high mass transfer rates, the
transition front can reach regions in the disk where
self-gravitation becomes important, and there the assumption of a
homogeneous disk becomes quite questionable, as a result of the
development of a gravitational instability that is likely to result
in the fragmentation of the disk.

\end{document}